# Role of contact work function, back surface field and conduction band offset in CZTS solar cell

Atul Kumar[*], Ajay D. Thakur

*Department of Physics, Indian Institute of Technology Patna, Bihta, India 801106*

[*]E-mail: atul.pph14@iitp.ac.in

We employ simulation based approach for enhancing the efficiency of $Cu_2ZnSnS_4$ (CZTS) based solar cells. Initial benchmarking of simulation with the experimentally reported solar cell in literature is performed by incorporating a suitable defect model. We then explore the effects of: (a) conduction band offset (CBO) at CZTS/CdS junction, (b) back surface field (BSF) due to an additional layer with higher carrier density, and (c) high work function back contact. Efficiency is observed to improve by about 70% upon optimizing the above three parameters. We also observe that utilizing BSF in the configuration can reduce the high work function requirement of the back contact. A work function of 5.2 eV (e.g., using Ni), a BSF layer (e.g., using SnS), and a CBO of 0.1 eV (e.g., using ZnS) constitute an optimal configuration.

# 1. Introduction

There is a continuing increase in energy demand despite the looming issues related to global warming, climate change and the limited fossil fuel resources. Solar energy being economical, eco-friendly and perennial source is therefore indispensable for meeting this demand. $Cu_2ZnSnS_4$ (CZTS) based solar cells where, all the ingredient elements are non-toxic and naturally abundant [1], with an economic fabrication process presents itself as a possible option for large scale deployment [2]. Wadia *et al* [1] estimated the price/Watt tag of CZTS to be about $4.9 \times 10^{-3}$ Cent/Watt. It should be noted here that CZTS does not suffer from the issue of Indium availability as is the case with CIGS based solar cells. CZTS solar cells belong to the category of photovoltaic devices with smallest $/Watt tag [1]. From the material's availability perspective, ingredient elements for CZTS ensure terawatt scale deployment at current efficiency of 9.2% [3-4]. The maximum reported laboratory scale efficiency is 12.6% [5] for CZTSSe (where a fraction of S atoms is replaced by Se atoms). The maximum efficiency was reported in 2013 and since then, enormous efforts are underway to further improve its performance. The key problems with CZTS based solar cell include: i) sub-optimal absorber layer, ii) Schottky barrier at back contact, and iii) recombination at front contact [6-7]. CZTS as bulk absorber suffers from a high density of vacancies, interstitial and antisite defects and secondary phases. It also has band gap tail states of multivalent defects near the conduction band (CB) and the valence band (VB) [8-9]. Performance parameters of various CZTS based solar cells reported in literature has been compiled to ascertain their performance compared to theoretical SQ limit and these are shown in Fig. 1 [10-22]. It can be noted that there is a significant deficit of the experimentally observed performance parameters compared to the SQ limit (solid lines in Fig. 1). A non-optimal absorber shows poor electrical properties (including low mobility, carrier lifetime related issues, high sheet resistance, band gap fluctuations and high density of defects) and phase purity issues (presence of binary phases and defects). A comparison of various physical properties of CIGS, CZTS and CZTSSe (where, a fraction of S atoms are replaced by Se atoms) solar cell materials is shown in Table I. From the Table I, it can be observed that the numerical values of various physical parameters for CZTS are very different from those of CIGS. These deviations are primarily responsible for a lower efficiency of CZTS based solar cells when compared to CIGS based solar cells. Despite these limitations, there

is a possibility of enhancing the efficiency of CZTS based solar cells using suitable optimization of Mo back and CdS front contact [7].

The unstable back contact at the Mo/CZTS junction causes secondary phase formation of $MoS_2$ causing a Schottky barrier [25]. Front contact with CdS has a lattice mismatch, enhanced interface recombination and an unfavorable type II conduction band offset (CBO) at CdS/CZTS junction [26]. The diode quality of CZTS/CdS is suboptimal with a high diode ideality factor [5] and a higher dark saturation current (~0.2 μA) than CIGS (about 0.1nA). These challenges can be addressed via suitable optimization of the solar cell architecture. Table II provides a guideline for important design issue in p-n heterojunction [27-28] and summary of earlier efforts on thin film solar cell optimization [30-44]. This includes optimization of (a) the ratio of band gap, carrier densities in the absorber and the buffer layers for reduced interfacial recombination [30], (b) back contact work function [31, 32, 35], (c) CBO and the band alignment at the junction [34], (d) band gap grading in $E_v$ and $E_c$ [33,36,39], (e) back surface field (BSF) [43-45]. A simulation of CZTS solar cell by SCAPS and optimization of back contact work function, acceptor concentration and thickness of CZTS layer was performed by Patel *et. al.* [32], Meher *et. al.* [40] simulated and studied the effect of thickness, carrier concentration, defect density, mobility, conduction band offset, capacitance-voltage curve and thermal admittance spectra on CZTS/CdS junction using SCAPS. Courel *et. al.* [41] simulated the effect of high minority carrier lifetime and low interface recombination rate owing to improved crystallinity for simulating structure with an efficiency of 18%. However none of these past efforts [30-44] take into account the effect of multivalent defects, which otherwise is considered to be a detrimental factor affecting the behavior of CZTS as a photovoltaic material [8-9]. Therefore, prior to incorporating any of the above modification in the standard solar cell configuration, it is worthwhile to incorporate a suitable defect model so that the simulated value of efficiency matches well with the experimentally observed numbers. We make use of such a defect model (described in the next section). We then perform a modeling based study of CZTS solar cell for performance enhancement focusing on: (a) merits of a high work function back contact, (b) implication of back surface passivation (BSP) layer which leads to an effective back surface field (BSF), (c) improvements in solar cell performance due to a spike-type band alignment at the absorber-buffer interface, and (d) the role of conduction band offset at the buffer-absorber and buffer-window heterojunctions. The results are discussed below.

## 2. Simulation Method and Model Details

There are number of simulator available which solve semiconductor and electrostatic equation in1D environment to simulate a thin film solar cell. Most of them (SCAPS, AMPS, PC1D, GPVDM) solves 1D Poisson and continuity equation for hetero-structures. We used the solar cell capacitance simulator SCAPS-1D version 3.0.2 [46-51] which solves the semiconductor and electrostatic equation of transport and continuity equation using Newton Ralphson and Gummel iteration. It performs 1D modeling of thin films polycrystalline solar cells like CIGS, CdTe, CZTS etc. It solves these equation for calculating the I(V), C(V),C(f), Quantum efficiency curve, band diagram, recombination in illumination in p-n heterojunctions structures. It can model the various tunneling and multivalent defect generally relevant in these thin film solar cell. Thus the effects of absorption, energy band alignments, inter and intra band tunneling, defects energy level, recombination can also be suitably illustrated by this software package in thin film solar cells.

Initially, a benchmarking is done via incorporation of a suitable defect model to match the simulation results of the baseline solar cell configuration with the corresponding experimentally reported values of Shin *et.al.* [22] We have chosen Shin *et.al.*[22] for our baseline cell as it is one of the highest efficiency reported for CZTS and the authors have provided all the relevant details of the CZTS device in their paper. To begin with, we have modeled an identical structure Mo/CZTS/CdS/ZnO/AZO and simulated the performance of the solar cell. In addition, for simulation purpose, we have used the reported material parameters of CZTS which include an absorber layer thickness of 600nm, band gap 1.45eV, dielectric constant of 7, an acceptor density $N_A$ of $\sim 10^{16}/cm^3$, diffusion length ($L_n$) for electrons in CZTS $\sim$ 350 nm, and an electron mobility of $\sim$5 cm$^2$/Vs. We have assumed multivalent defects with charge states {3+/2+, 2+/+, +/0} in CB located at defect energy level, $E_t$=1.40eV with characteristic energy of 0.05eV. Acceptor defects of charge state {0/-, -/2-, 2-/3-} at VB. CZTS has a wide variety of point defects which include: (i) vacancies ($V_{Cu}$, $V_{Zn}$, $V_{Sn}$ and $V_S$), (ii) interstitials ($Cu_i$, $Zn_i$, $Sn_i$ and $S_i$), (iii) antisites ($Cu_{Zn}$, $Zn_{Cu}$, $Zn_{Sn}$, $Sn_{Zn}$, $Cu_{Sn}$) etc. These defects can also form defect clusters, e.g., $V_{Cu}+Zn_{Cu}$, $2Cu_{Zn}+Sn_{Zn}$, $Zn_{Sn}+2Zn_{Cu}$, etc. [8-9]. Point defects and their details are summarized in Table

III. SCAPS has been used to simulate multivalent defect occurring in chalcopyrite and chalcogenide solar cell materials [52]. We have utilized multivalent defects in the bulk of CZTS absorber layer to match our baseline solar cell to the experimentally reported CZTS solar cell [22]. We used a metal work function of 5eV. The surface recombination velocity (SRV) used at left contact are $10^5$ cm/s (for electron) and $10^7$ cm/s (for hole). The SRV used for right contact are $10^5$ cm/s (holes) and $10^7$cm/s (electron). SRV can be approximated by SRV = $N_{st}\rho V_{th}$, where, $V_{th}$ is thermal velocity ($10^7$ cm/s), $\rho$ is the capture cross section for charge carriers and $N_{st}$ is surface defect density. For an n-type semiconductor/metal contact and p-type semiconductor/metal contact, SRV will depend upon the interface defect states and the capture cross sections of electron and hole. Our values are higher as we have not used any passivation in the benchmarking. Similar values are used by Patel *et. al.* [32] and Meher *et. al.*[40]. For a passivated contact SRV values will be lower and correspondingly performances will be higher [53] and therefore our choices do not over-estimate the efficiency. The additional model inputs include: (a) a defect density of $5\times10^{14}$ cm$^{-3}$, (b) a capture cross section of $10^{-13}$cm$^2$, (c) a radiative recombination coefficient of $10^{-13}$cm³/s, and (d) an Auger capture coefficient of $10^{-28}$cm$^6$/s. Auger recombination becomes relevant when the carrier density increases. The Auger recombination coefficient is approximately $10^{-30}$cm$^6$/s in Si and $10^{-27}$cm$^6$/s in GaAs. We have used an Auger coefficient value of $10^{-28}$ cm$^6$/s in CZTS. Typically, Auger coefficient does not have any significant effect on performance until it is higher than$10^{-25}$ cm$^6$/s at a doping density of $10^{16}$cm$^{-3}$. For CZTS, an optimized Auger capture coefficient lies in the range $10^{-29}$-$10^{-28}$cm$^6$/s [32]. At the CZTS/CdS interface, a neutral defect density of $10^{12}$ cm$^{-3}$ is taken. We have also made use of the inbuilt absorption file for absorption coefficient of the buffer layer (CdS.abs) which in turn is based on Adachi [54]. For the ZnO and AZO layers, we have used ZnO.abs file as given in SCAPS version 3.3.02 [54]. It should be noted that although the absorption coefficient data for AZO has been taken to be the same as ZnO, however, the donor/ acceptor density ($N_D/N_A$) has been taken as different from ZnO. Also, a slight blue shift has been done to the electron affinity value for AZO compared to ZnO to implement a type-II band (cliff-type) alignment [34]. The device is illuminated from window layer side with AM1.5 spectrum using relevant SCAPS option at temperature 300 K. The relevant material parameters used in simulation are listed in Table IV [32, 39-41]. The simulated current-voltage (J-V) curve in main panel of Fig.2 shows a $V_{oc}$ of 0.66 V, a $J_{sc}$ of 19.85mA/cm$^2$, a FF of 63.6% and an efficiency of 8.42%. The quantum efficiency (QE) curve is also matching with the experimental QE curve [22] with maximum absorption in CZTS region. The model parameters used along with the chosen defect model leads to a good agreement of our simulation results with the experimental data obtained by Shin *et.al.* [22]. The parameter and the defect model is kept identical across the rest of the simulations involving modifications to the original solar cell configurations.

## 3. Results and discussion

### 3.1 Description of p-n heterojunction solar cell

The band alignment at conduction band of the absorber and buffer is crucial. Interfacial recombination taking place at junction between electrons from buffer ($E_{C;\ buffer}$) layer recombining with the holes of absorber in valence band ($E_{V;abs}$) labeled by 2 in Fig. 3. The reduced energy difference between absorber ($E_{V;abs}$) and buffer ($E_{C;\ buffer}$) will increase the recombination. A type1 heterojunction band alignment at $E_{CBO}$ (as shown in Fig. 3) can reduce the chance of interfacial recombination. The $E_{CBO}$ (as shown in Fig.3) acts as the Schottky barrier for photo generated carriers in the absorber layer. This blocking effect for type-I junction is sub-dominant until $E_{CBO}$ is less than the built-in voltage ($V_{bi}$) in the absorber layer. Thus, a slightly positive $E_{CBO}$ can reduce the recombination and is crucial for obtaining optimum values of $V_{oc}$ and FF.

The collection of generated and free carriers can be enhanced by applying electric field and providing addition drift to carriers at appropriated place. The presence of an electric field can be setup using a high-low junction [45]. Back surface field (BSF) can be setup at the rear by utilizing p$^+$-p high-low junction it will provides a drift to holes and block electron at the back contact. The n-n$^+$ high-low junction at the front provide an additional drift to electrons and enhance their collection probabilities. An additional advantage of high low junction at back and front is that the contacts acquire a near-ohmic character.

At the back contact, a Schottky barrier forms if $\phi_m < E_g + (\delta_{Abs} + \phi_d)$ where, $\phi_m$ is the work function of metal, $E_v$ is the valence band energy, $E_g$ is band gap, $\delta_{Abs}$ is electron affinity of absorber material, $\phi_d$ is the energy difference in the valence (or conduction) band in bulk and at the edge. Fig. 3 depicts the schematic of the p-n$^+$-n$^{++}$ showing back contact marking all relevant energy levels. The Schottky barrier ($\phi_s$) is given by $\phi_s = \phi_m - E_V$ at the metal contact, The reduction in the Schottky barrier height approaches a near ohmic nature. For a given absorber material, the variation in Schottky barrier height can be obtained by choosing a back contact of work function close to ($\delta_{Abs} + \phi_d + E_g$). The bending of the band at the back contact is due to the surface defect states and high surface recombination velocities. Corresponding Fermi level pinning and high Schottky barrier lowers the $V_{bi}$ in a typical device. An optimal configuration of heterojunction would thus comprise of a p-n$^+$ type junction. The p absorber layer should have an optimum band gap (1.4 eV) and the n layer has a high band gap. There will be a slight positive CBO at the junction (spike-type band alignment). A near-ohmic contact is achieved by utilizing a suitable high work function metal. Suitably placed electric field for high collection efficiency can be achieved by using additional layers.

## 3.2 Role of high work-function back contact

Mo as a back contact in CZTS solar cells reacts with sulphur in CZTS and forms a thin layer of $MoS_2$ [55]. The Mo/CZTS junction acts as a blocking back contact [21]. Schottky contact of p-type semiconductor CZTS with Mo as back contact is sub-optimum. It presents a barrier for majority carrier crossover. At the back contact, a Schottky barrier forms if $\phi_m < E_g + (\delta_{Abs} + \phi_d)$ where, $\phi_m$ is the work function of metal, $E_v$ is the valence band energy, $E_g$ is band gap, $\delta_{Abs}$ is electron affinity of absorber material, $\phi_d$ is the energy difference in the valence (or conduction) band in bulk and at the edge. Fig. 3 depicts the schematic of the p-n$^+$-n$^{++}$ showing back contact marking all relevant energy levels. The Schottky barrier ($\phi_s$) is given by $\phi_s = \phi_m - E_V$ at the metal contact. The reduction in the Schottky barrier height approaches a near ohmic nature. For a given absorber material, the variation in Schottky barrier height can be obtained by choosing a back contact of work function close to ($\delta_{Abs} + \phi_d + E_g$). Alternatively, a metal contact is made with a high work function metal ($\phi_m$) in case of p$^+$ absorber layer [31, 56-57]. In this context, we have simulated the CZTS solar cell with varying work function.

Simulation is performed for work function values in the range 4.6-5.65eV. The quantity ($\phi_m - E_V$) is the mismatch at back junction which is negative and becomes zero at high $\phi_m$ (of 5.65eV). In the main panel of Fig.4, the band diagram is shown for different $\phi_m$ values.

Variation of $V_{bi}$ is plotted as a function of metal work function in the inset panel of Fig.4. The band energy level in CZTS shifts upwards with increasing $\phi_m$. $V_{bi}$ grow linearly with $\phi_m$ till 5.3eV and remains almost constant after that. This built in voltage is the net potential drop across the absorber width. This potential drop is responsible for setting an electric field in the direction of p-n junction. In band-diagram plot, the energy levels in absorber bend downwards at the metal contact. This is due to surface defect states and the high SRVs used in the simulations. Bending is absent for SRV value of 10cm/s. Close to flat band condition this downward bending changed the slope. The overall potential is unidirectional setting a favorable unidirectional flow across the absorber layer. The 1$^{st}$ derivative of $E_c$ (or $E_v$) corresponds to the electric field. The presence of such potential/field is favorable for majority drift within the absorber layer towards the junction. Such favorable alignment can be employed in solar cell structure at appropriate locations to induce directional flow of charge carriers in solar cells for selective contacts. The electric field in the direction of p-n junction will drift the generated carriers towards the junction, aiding the diffusion length of the carriers and higher carrier collection at the n side. The high work function contact at back enhances the built-in voltage of the device [56].

Fig.5 shows the effect of back contact modification on performance parameters. A normalized plot of $V_{oc}$, $J_{sc}$, FF and efficiency is shown as a function of the metal work function in the inset of Fig. 5. These parameters are normalized by dividing the quantities with their corresponding highest values in the given range. While, $V_{oc}$ is highly sensitive to $\phi_m$ values, $J_{sc}$ and FF has a very small dependence on $\phi_m$. Efficiency follows the trend for $V_{oc}$. The efficiency of the device as a function of the back contact work function is shown in the inset of Fig.5. An efficiency of 13.8% is obtained with a work function of 5.6eV for the back contact. The efficiency is seen to improve by increasing $\phi_m$ for $\phi_m \leq E_V$. Simulations do not converge for the $\phi_m > E_V$. Modification of work function at back contact therefore has the potential for increasing the efficiency. Optimized value of Work function in our case is 5.6eV which is very high. Some of the materials with a high work function which can be applied as a back contact include Ni, Pt, Se and Au. An alternative technique needs to be evolved for lowering high work function which is discussed in next section.

### 3.3 Role of back surface field (BSF)

When a junction is made between a p type CZTS (charge density$\approx 10^{16}$) absorber layer and the n type CdS (charge density $10^{17}$) buffer layer, the space charge width will be shifted into the CZTS layer due to a p-n$^+$ type structure. Due to this, a combined drift and diffusion carrier transport comes into picture leading to enhanced collection of photogenerated carriers. Similarly an electric field at back of CZTS can be induced with the help of a high-low junction. A p$^+$ layer at back will make a p$^+$-p high-low junction and create an effective field (which is BSF) in direction of the p-n$^+$ junction [44] at CZTS/CdS. The effect of such a modification (p$^+$-p-n$^+$) will be a selective contact at back of absorber layer and an improved collection of photogenerated carriers. The inclusion of p$^+$-p will also make a near ohmic contact with metal at the back, similar to the front surface field arising from the nn$^+$ buffer-window (CdS/ZnO/AZO) junction. We have included a layer of hole doping $10^{18}$ cm$^{-3}$ and similar property as absorber layer of the BSF layer.

We observe that the efficiency increases to 11.3% in the presence of the resulting back surface field. The rise in V$_{OC}$ with BSF is caused by higher built-in field across absorber. The band diagram and IV curve are shown in the Fig. 6. The IV curve shows a roll over effect in Fig.6 (b) which is due to the back contact barrier as shown in the band diagram of Fig 6 (b). The p$^+$ layer has a higher barrier with Mo contact at the back than the p type CZTS. We performed the back contact work function optimization for BSP layer to make an ohmic contact. The barrier decreases with the increasing work function as shown in Fig. 6 (c). Band diagram in Fig.6 (d) shows an absence of energy level bending at the back. It is observed that BSP layer alone leads to a smaller increment in the efficiency but it reduces the optimum work function of back contact. Figure 6 (c) shows that the optimized work function obtained in conjunction with a BSP layer is 5.2 eV. This value is less than the optimized work function 5.6eV in section 3.3.1. The final efficiency is 13.8% in both cases of back contact work function optimization (in section 3.1) and a combined BSP layer with work function optimization. Metals like Ni, Au ($\phi_m$~5.2eV) can be used as a back contact along with a p$^+$ BSP layer of a material like SnS for an optimum device structure.

## 3.4 Front Contact: Spike-type band alignment

Higher efficiency CIGS/CdS has a type-I interface with a spike-type junction [58]. Similar band offset for CZTS/CdS may result in a higher efficiency. The alignment of the energy level controls the charge flow through the CZTS/CdS junction. Any mismatch will give rise to a band offset. Fig. 3 shows the schematic of the band alignment. Conduction band offset ($E_{CBO}$) at the junction is given by $\delta_{CZTS}-\delta_{CdS}$ where $\delta$ is the electron affinity. When $\delta_{CZTS}>\delta_{CdS}$, $E_{CBO}$ is positive and a barrier (spike-type) will be present for photo-generated electron flow from absorber to buffer layer. This configuration at the junction is a type-I heterojunction or straddling gap. $E_{CBO}$ is negative when $\delta_{CZTS}<\delta_{CdS}$ and this corresponds to a type-II band gap and a cliff will be present at the junction. In the present case of CZTS/CdS junction, a band bending is occurring at the junction. This is given by $V_{bi}$ as shown in Fig. 3. Changes due to the band bending in $E_{CBO}$ is $(\delta_{CZTS}+V_{bi}-\delta_{CdS})$. CZTS/CdS has a type-II heterojunction as reported experimentally [22]. Band offset in the CZTS/CdS/ZnO solar cell at the junctions CZTS/CdS ($CBO_{CZTS/CdS}$) and CdS/ZnO ($CBO_{CdS/ZnO}$) are implemented in the simulation. Main panel (and the inset) of Fig.7 depict the ($CBO_{CZTS/CdS}$), ($CBO_{CdS/ZnO}$) and $\phi_b^n$ (energy barrier for electron from $E_{Fn}$ of absorber to maximum of buffer $E_C$). $CBO_{CZTS/CdS}$ values for the case of positive, negative and zero conduction band offset is simulated and shown. Electron affinity of the CdS ($\delta_{CdS}$) (for the fixed value of $\delta_{CZTS}$ of 4.2eV) is varied to change CBO. Cliff at junction of -0.1eV (negative CBO) is shown by the dot-dashed line and the dashed line correspond to a spike of 0.1eV in Fig. 7. An electron affinity lower than that of CZTS is used for the buffer layer leading to a type-I interface with $CBO_{CZTS/CdS}>0$. Apart from band alignment, interface defects arising from lattice mismatch and thermal expansion coefficient differences at the CZTS/CdS junction leads to reduction in current. Simulation results show that the recombination current is lowest in positive conduction band offset region $CBO_{CZTS/CdS}>0$. Slightly positive conduction band offset shows lowest recombination current and maximum efficiency as shown in Fig. 8. At $CBO_{CZTS/CdS}=0.05eV$, the efficiency deteriorates at only highly negative $CBO_{CdS/ZnO}$ values and remains insensitive at other values as shown in inset of Fig. 7. Simulation suggests that a slightly positive $CBO_{CZTS/CdS}$ and $CBO_{CdS/ZnO}>-0.3eV$ are optimum conditions for a high efficiency. For the value of $\phi_b^n$ $<V_{bi}$, the band alignment will not present a barrier for photo-generated carriers [38]. Inset panel of Fig.7 shows normalized performance parameter as a function of $CBO_{CZTS/CdS}$.

Normalization is done by dividing the value by maximum value thus setting highest value to one. Fill factor is sensitive to CBO at the absorber/buffer junction as sheen in Fig. 8. Type-I heterojunction with slightly positive CBO$_{CZTS/CdS}$ is optimum for the present case. Similar behavior was seen in other systems [38,59]. For practical applications, ZnS could be such a layer fulfilling the above criteria for practical purpose. It has similar structure as CZTS and a wide band gap. Such a modification has the potential to improve the otherwise poor ideality factor and a high diode saturation current in CZTS based solar cell.

## 3.5 Overall device performance

Optimization of both the front and back contacts is performed in the simulation. The improvements observed upon incorporating a BSF layer, a high work function back contact and a spike-type junction include: (i) increase in Voc with BSF layer and a high work function metal back contact (=5.2 eV), (ii) reduction in recombination at the CZTS/buffer junction which in turn increases the FF and the overall efficiency. The standard structure Mo/CZTS/CdS/ZnO/AZO has an efficiency of 8.4%. The inclusion of a high work function metal back contact alone optimized efficiency to 13.8%, however, the work function required is higher (=5.6 eV). The inclusion of BSF layer alone optimized efficiency from 8.4 to 11.3%. The inclusion of a positive CBO (spike-type) alone at CZTS/CdS junction optimized efficiency from 8.4 to 8.75%. If we consider the following two scenarios: (A) high work function metal and a spike-type band alignment at front contact, and (B) high work function metal, a BSF layer and a spike-type band alignment at front, then the efficiencies are nearly same ($\approx$14.4%). In case (A), the optimum work function is 5.6 eV whereas in case (B) it is 5.2eV, which is more realistic. The standard and modified configuration of high work function at back and spike-type band structure at CZTS/CdS junction is shown in the Fig.9. Band diagram of the two solar cell configurations show the energy variation and the width of the different layers. The simulation output with these modifications is compared with the simulation results for the standard CZTS based solar cell configuration. Improvement in quantum efficiency and the current-voltage (J-V) characteristics is shown in Fig.10. The observed performance parameters in the optimized solar cell configuration are V$_{oc}$= 0.921V, J$_{sc}$= 21.88mA/cm$^2$, FF =71.6% and an efficiency of 14.4% in the modified structure. There is aproximately 70% improvement in CZTS performance from 8.4% to 14.4% after incorporating these changes.

## 4. Conclusions

We benchmarked our simulations against one of the experimentally reported CZTS solar cell after incorporating suitable multivalent defect model. Although CZTS/CdS forms a typical p-n$^+$ type heterojunction, its band alignment is unoptimised. Performance optimization is achieved by introduction of suitable back contact, BSF layer and a spike-type band alignment at the front contact. Our simulations reveal that a high work function contact at the back and a spike-type band alignment at the front contact are required for higher performance of the device. An additional BSF layer can be used to lower the high work function back contact requirement without lowering the performance. A slightly positive CBO making a spike-type junction at the CZTS/CdS is advantageous for improving performance. For a spike-type band alignment at conduction band of CZTS/buffer junction, ZnS could be a replacement for CdS and this also eliminates toxicity concern of CdS. With these modifications, the efficiency is predicted to increase from 8.4 % to 14.4% leading to an overall improvement by about 70%. The work function of back contact around 5.2eV (e.g., Ni), a BSP layer (e.g., SnS), an absorber (CZTS) and a buffer layer (e.g., ZnS) with CBO of 0.1eV could therefore constitute the desired architecture.

## Acknowledgments

We gratefully acknowledge Dr Marc Burgelman, Honorary Professor, University of Gent for providing SCAPS-1D software.

## References

1) C. Wadia, A. P. Alivisatos, and D. M. Kammen, Environ. Sci. Technol. **43**, 2072 (2009).

2) Y.E. Romanyuk, C.M. Fella, A.R. Uhl, M. Werner, A.N. Tiwari, T. Schnabel, E. Ahlswede, Solar Energy Materials and Solar Cells **119**, 181 (2013).

3) K. Sun, C. Yan, F. Liu, J. Huang, F. Zhou, J. A. Stride, M. Green, X. Hao, Adv. Energy Mater. **6**, 1600046 (2016).

4) D. B. Mitzi, O. Gunawan, T.K. Todorov and D. A. R. Barkhouse, , Phil. Trans. R. Soc. A **371**, 20110432 (2013).

5) W. Wang, M. T. Winkler, O. Gunawan, T. Gokmen, T.K. Todorov, Y. Zhu, and D. B. Mitzi, Adv. Energy Mater. **4**, 1301465 (2014).

6) T. Gershon, T. Gokmen, O. Gunawan, R. Haight, S. Guha, and B. Shin, MRS Communications **4**, 159 (2014).

7) A. Polizzotti, I. L. Repins, R. Noufi, S. Wei, and D. B. Mitzi, Energy Environ. Sci. **6**, 3171 (2013).

8) S. Chen, A. Walsh, X. Gong, and S. Wei, Adv. Mater. **25**, 1522 (2013); A. Walsh, S Chen, S. Wei, and X.

Gong, Adv. Energy Mater. **2**, 400 (2012).

9) S. Chen, L.Wang, A.Walsh, X. G. Gong, and S.Wei, Appl. Phys. Lett. **101**, 223901 (2012).

10) S. Rühle, Solar Energy, Solar Energy **130**, 139 (2016).

11) K. F. Tai, O. Gunawan, M. Kuwahara, S. Chen, S. G. Mhaisalkar, C. H. A. Huan, and D. B. Mitzi, Adv. Energy Mater. **6**, 1501609 (2016).

12) S. Bourdais, C. Choné, B. Delatouche, A. Jacob, G. Larramona, C. Moisan, A. Lafond, F. Donatini, G. Rey, S. Siebentritt, A. Walsh, and G. Dennler, Adv. Energy Mater. **6**, 1502276 (2016).

13) C. K. Miskin, W.Yang, C. J. Hages, N. J. Carter, C. S. Joglekar, E. A. Stach and R. Agrawal, Prog. Photovolt: Res. Appl. **23**, 654 (2015).

14) T. K. Todorov, K. B. Reuter, and D. B. Mitzi, Adv. Mater. **22**, 156 (2010).

15) Z. Tong, K. Zhang, K. Sun, C. Yan, F. Liu, L. Jiang, Y. Lai, X. Hao, J. Li,    Solar Energy Materials & Solar Cells **144**, 537 (2016).

16) S. G. Haass, M. Diethelm, M. Werner, B. Bissig, Y. E. Romanyuk and A. N. Tiwari, Adv. Energy Mater. **5**, 1500712 (2015).

17) H. Hiroi, N. Sakai, T. Kato, and H. Sugimoto, IEEE 39th Photovoltaic Specialists Conference (PVSC) (2013).

18) W. Yang, H. Duan, B. Bob, H. Zhou, B. Lei, C.Chung, S. Li, W. W. Hou, and Y. Yang, Adv. Mater. **24**, 6323 (2012).

19) G. Larramona, S. Bourdais, A. Jacob, C.   Choné, T. Muto, Y. Cuccaro, B. Delatouche, C. Moisan, D. Pe´re´, and G. Dennler, J. Phys. Chem. Lett. **5**, 3763 (2014).

20) I. L. Repins, M. J. Romero, J. V. Li, S. Wei, D. Kuciauskas, C.Jiang, C. Beall, C. DeHart, J. Mann, W. Hsu, G. Teeter, A. Goodrich, and R. Noufi, IEEE Journal of Photovoltaics **3**, 439 (2013).

21) T. Gokmen, O. Gunawan, and D. B. Mitzi, Journal of Applied Physics **114**, 114511 (2013).

22) B. Shin, O. Gunawan, Y. Zhu, N. A. Bojarczuk, S. J. Chey, and S. Guha, Prog. Photovolt: Res. Appl. **21**, 72 (2013).

23) P. Jackson, D. Hariskos, R. Wuerz, O. Kiowski, A. Bauer, T. M. Friedlmeier, and M. Powalla, Phys. Status Solidi RRL **9**, 28 (2015).

24) S. A. Dinca and E. A. Schiff, B. Egaas, R. Noufi, D. L. Young, and W. N. Shafarman, Phys. Rev. B **80**, 235201 (2009).

25) O. Gunawan, T. K. Todorov, and D. B. Mitzi, Appl. Phys. Lett.**97**, 233506 (2010).

26) Z. Dong, Y. Li, B Yao, Z. Ding, G. Yang, R. Deng, X. Fang, Z. Wei and L. Liu,    J. Phys. D: Appl. Phys. **47**, 075304 (2014).

27) U. Rau, H.W. Schock, Appl. Phys. A **69**, 131(1999).

28) A. Niemegeers, M. Burgelman, and A.D. Vos, Appl. Phys. Lett. **67**, 843 (1995).

29) H J Pauwels and G Vanhoutte . J. Phys. D: Appl. Phys., **11**, 649 (1978).

30) R. Klenk, Thin Solid Films **387**, 135, (2001).

31) S.H. Demtsu, J.R. Sites, Thin Solid Films **510**, 320 (2006).

32) M Patel, A. Ray, Physica B, **407** 4391(2012).

33) Y. Hirai, Y. Kurokawa, and A. Yamada, Japanese Journal of Applied Physics **53**, 012301 (2014).

34) G. Sozzi, F. Troni, R. Menozzi, Solar Energy Materials & Solar Cells **121**, 126 (2014).

35) T. Minemoto, M. Murata, Current Applied Physics **14**, 1428 (2014).

36) D. Hironiwa, M. Murata, N. Ashida, Z. Tang, and T. Minemoto, Japanese Journal of Applied Physics **53**, 071201 (2014).

37) C. Huang, W. Chuang, Vacuum **118**, 32 (2015).

38) T. Minemoto, M. Murata, Solar Energy Materials & Solar Cells **133**, 8 (2015).

39) O.K. Simya, A. Mahaboobbatcha, K. Balachander, Superlattices and Microstructures **92**, 285 (2016).

40) S.R. Meher, L. Balakrishnan, Z.C. Alex, Superlattices and Microstructures **100**, 703 (2016).

41) M. Courel, J A Andrade-Arvizu, and O Vigil-Galánl, Mater. Res. Express **3**, 095501,(2016).

42) D. Cozza, C. M. Ruiz, D. Duch, J. J. Simon, and L. Escoubas, IEEE Journal of Photovoltaics. **6**, 5 (2016).

43) M. K. Omrani, M. Minbashi, N. Memarian, D. Kim, Solid State Electronics **141**, 50 (2018).

44) A. Kumar and A. D. Thakur, https://arxiv.org/ftp/arxiv/papers/1510/1510.05092.pdf.

45) O. von Roos, Journal of Applied Physics **49**, 3503 (1978).

46) M. Burgelman, K. Decock, S. Khelifi, A. Abass, Thin Solid Films **535**, 296 (2013).

47) S. Khelifi, M. Burgelman, J. Verschraegen, A. Belghachi, Solar Energy Materials & Solar Cells **92**,1559 (2008).

48) M. Gloeckler, J.R. Sites, Thin Solid Films **480**, 241 (2005).

49) M. Boumaour, S. Sali, A. Bahfir,S. Kermadi, L. Zougar, N. Ouarab, and A. Larabi, J. Electron. Mater. **45**, 8 (2016).

50) A. Chihi, M.F. Boujmil, and B. Bessais, J. Electron. Mater., **46**, 8 (2017).

51) M . Houshmand, M.H. Zandi, and N.E. Gorji, JOM **67**, 2062 (2015).

52) K. Decock, S. Khelifi, M. Burgelman, Thin Solid Films **519**, 7481 (2011).

53) B.Vermang, V. Fjällström, J. Pettersson, P. Salomé, M. Edoff, Solar Energy Materials & Solar Cells **117,** 505 (2013).

54) S. Adachi*, Optical constants of crystalline and amorphous semiconductors.* (Springer, Boston, 1999) p.426-42

55) J. J. Scragg, J. T. Wätjen, M. Edoff, T. Ericson, T. Kubart, and C. Platzer-Bjorkman, J. Am. Chem. Soc. **134,** 19330 (2012).

56) T. Jäger, Y. E. Romanyuk, B. Bissig, F. Pianezzi, S. Nishiwaki, P. Reinhard, J. Steinhauser, J. Schwenk, and A. N. Tiwari, Journal of Applied Physics **117**, 225303 (2015).

57) D. Rached, R. Mostefaoui, Thin Solid Films **516**, 5087 (2008).

58) S. Huang, W.Luo,and Z. Zou,   J. Phys. D: Appl. Phys. **46**, 235108 (2013).

59) R. Scheer, Journal of Applied Physics **105**, 104505 (2009).

## Figure Captions

**Fig. 1.** (Color online) Plot of solar cell performance parameters as a function of band gap for laboratory solar cells reported in the literature (filled circles), guide to eye for the variation with $E_g$ (dot-dashed line) and the corresponding SQ limit (solid line) for: (a) $J_{SC}$, (b) $V_{OC}$, and (c) FF. The double sided arrows mark the deficit in the observed solar cell parameters (reported in literature [10-21]) compared to the corresponding SQ limit.

**Fig. 2** Plot of simulated current-voltage (J-V) characteristics (main panel) and quantum efficiency (QE) versus wavelength (inset panel) for the standard CZTS solar cell with suitable defect model (see text) [22].

**Fig. 3.** (Color online) The band alignment of p-n$^+$-n$^{++}$ configuration of thin films solar cell with spike-type (type-I) band alignment. Various level spacing are marked using suitable arrows (see text).

**Fig. 4.** Main panel: Band diagram of the CZTS/CdS/ZnO/AZO configuration for different metal work function ($\phi_m$) values. Inset: Variation of built-in voltage $V_{bi}$ with $\phi_m$.

**Fig. 5.** Main panel: Normalized performance parameters ($V_{OC}$, $J_{SC}$, FF, and Efficiency) as a function of $\phi_m$. Inset: Efficiency versus $\phi_m$.

**Fig. 6.** Shows the band diagram after including the BSP layer. (b) The J-V curve of the cell with and without BSP layer. It shows a roll over which arises due to the contact barrier shown in (a). (c) Efficiency of the solar cell containing BSP layer as a function of the back contact work function. (d) Band diagram of the solar cell incorporating the BSP layer at the optimized value (=5.2eV) of work function.

**Fig. 7.** Main panel: Simulated band alignment at CZTS/CdSjunction for conduction band offset at CZTS/CdS interface (CBO$_{CZTS/CdS}$) fixed at values of -0.1 eV, 0 eV and 0.1 eV, respectively. Inset: Zoomed out image of the simulated band alignment diagram across a larger region of the solar cell configuration.

**Fig. 8.** Main panel: Recombination current as a function of CBO$_{CZTS/CdS}$. Inset: Normalized performance parameters (V$_{OC}$, J$_{SC}$, FF,Efficiency) as a function of CBO$_{CZTS/CdS}$. Also shown in the inset is the variation of efficiency as a function of the conduction band offset at CdS/ZnO interface (CBO$_{CdS/ZnO}$) at a fixedvalue of CBO$_{CZTS/CdS}$(= 0.05 eV).

**Fig. 9.** Energy band diagram for: (a) standard CZTS based solar cell configuration and (b) modified configuration with a high work function back metal contact and a spike-type (type-I)alignment at CZTS/CdS junction.

**Fig. 10.** Plot of: (a) J-V characteristics, and (b) QE versus wavelength for the standard and modified CZTS based solar cell configurations.

## Table Captions

**Table I.** One to one comparison of CZTS and CIGS solar cell parameters.

**Table II.** The tabulation of guidelines for minimum interfacial recombination and enhanced collection efficiency in p-n junction device summarized from reported simulation work [30-44].

**Table III.** Point defects of the CZTS used for modeling the bulk defect [8-9].

**Table IV.** Material parameters used in simulation [32, 39-41].

**Table I**

| Device Parameters of Absorber layer | CZTS[22] | CZTSSe[5] | CIGS[23-24] |
|---|---|---|---|
| Minority carrier diffusion length ($L_d$) (μm) | 0.35 | 1 | ~2 |
| Minority carrier lifetime (ns) | ~5 | ~1-10 | ~10-100 |
| Minority carrier mobility (cm$^2$/Vs) | | 690 | 480-1600 |
| Hall mobility majority carrier (cm$^2$/Vs) | 6-30 | ~35 | ~1000 |
| $J_o$ saturation current (mA/cm$^2$) | ~$10^{-6}$ | $7 \times 10^{-8}$ | $2.4 \times 10^{-11}$ |
| Hole density (cm$^{-3}$) | ~$10^{16}$ | ~$10^{16}$ | $10^{16}$ |
| Efficiency (%) | ~8.4% | ~12.6% | ~21.7% |
| Band gap (eV) | 1.45 | 1.13 | 1.13 |
| Eg/q-Voc (V) | 0.77 | 0.62 | 0.42 |

**Table II**

| Issues to optimize | Guide line for optimum device | References | Simulator |
|---|---|---|---|
| Interfacial recombination | Asymmetric doping (p-n$^+$) [26] Asymmetric Band gap (E$_{g\ abs}$<E$_{g\ buffer}$) | Klenk optimized p-n$^+$ junction and CIGS/CdS solar cell [30] | SCAPS |
| | Band alignment (spike-type) [26-29] | Meher et. al. [40], Minemoto et.al. [38] emphasized band alignment | SCAPS |
| | | Sozzi et.al. optimized the abs./buffer band alignment in thin film solar cell [34] | Synopsys Sentaurus Mathemetica |
| | | Courel et. al. emphasized CZTS/CdS junction [41] | |
| Collection enhancement | Electric field (BSF , FSF) (p-n$^+$-n$^{++}$) [45] | Kumar et. al. [44], Omrani et.al. utilized BSF [43] | SCAPS |
| | Band gap profiling | Hirai et.al. [33], Hironiwa et.al. [36]. and Simya et. al. [39] have explored band profiling | SCAPS |
| Contact optimization | Work function of contacts [31, 37] | Patel et. al. [32], Minemoto et.al. [35] emphasized the detail of the work function of contact, D. Cozza et.al. [42]optimized Mo and MoSe$_2$ contact in CZTSSe solar cell | SCAPS |
| | Passivation for low Recombination velocities | | |

**Table III**

| CZTS | Defect type | Charge state | Approximate ionization levels of defects in bandgap |
|---|---|---|---|
| $V_{Cu}$ | Acceptor | (-/0) | 0.02 eV above VBM |
| $Cu_{Zn}$ | Acceptor | (-/0) | 0.15 eV above VBM |
| $Zn_{Sn}$ | Acceptor | (2-/0) | 0.2 eV above VBM |
| $Cu_{Sn}$ | Acceptor | (-/0)(2-/-)(3-/2-) | Multiple level |
| $V_{Zn}$ | Acceptor | (2-/0) | 0.2 eV above VBM |
| $V_{Sn}$ | Acceptor | (-/0)(3-/2-)(4-/3-) | Multiple level |
| $Zn_{Cu}$ | Donor | (0/+) | 1.4 eV above VBM |
| $Cu_i$ | Donor | (0/+) | 1.4 eV above VBM |
| $Sn_{Zn}$ | Donor | (0/+)(+/2+) | 1.4 eV above VBM |
| $Sn_{Cu}$ | Donor | (0/+)(+/2+)(2+/3+) | Multiple level |
| $Zn_i$ | Donor | (0/2+) | 1.4 eV above VBM |
| $Sn_i$ | Donor | (0/2+)(2+/3+)(3+/4+) | 1.4 eV above VBM |

VBM (Valence band maxima)

## Table IV

| Contact Properties | Front | Back |
|---|---|---|
| SRV electron (cm/s) | $10^7$ | $10^5$ |
| SRV hole (cm/s) | $10^5$ | $10^7$ |
| Work function ($\phi_m$) | flat band | varied (4.6-5.6eV) |
| Band offset (eV) | $E_{CBO}$ (CZTS/CdS) varied (-0.2 to 0.3) | |
| Radiative recombination coefficient (cm³/s)/Auger capture coefficient(cm⁶/s) | $10^{-15}/10^{-28}$ | |

| Parameters | CZTS | CdS | ZnO | AZO |
|---|---|---|---|---|
| Thickness (nm) | 600 | 80 | 80 | 450 |
| Band gap (eV) | 1.45 | 2.4 | 3.3 | 3.3 |
| Electron affinity (eV) | 4.2 | 4.25 | 4.5 | 4.52 |
| Dielectric permittivity | 7 | 9 | 9 | 9 |
| CB effective density of state $N_C$ (cm⁻³) $\times 10^{18}$ | 2.2 | 18 | 2.2 | 2.2 |
| VB effective density of state $N_V$ (cm⁻³) $\times 10^{18}$ | 18 | 2.4 | 18 | 18 |
| Electron thermal velocity (cm/s) | $10^7$ | $10^7$ | $10^7$ | $10^7$ |
| Hole thermal velocity (cm/s) | $10^7$ | $10^7$ | $10^7$ | $10^7$ |
| Electron mobility (cm²/Vs) | 5 | 50 | 30 | 30 |
| Hole mobility (cm²/Vs) | 0.275 | 5 | 5 | 5 |
| Donor/ Acceptor density, $N_D/N_A$ (cm⁻³) | $10^{16}$ | $10^{17}$ | $10^{18}$ | $10^{20}$ |
| Absorption coefficient | $8\times10^4$ cm⁻¹ | $9\times10^3$-$10^7$ m⁻¹ for wavelength520-272nm* | $1$-$10^7$ m⁻¹ for wavelength 425-206nm* | Same as ZnO |
| **Bulk defect[8-9]** | | | | |
| Multivalent defects | $N_t$ | Charge state; type | $\sigma_e;\sigma_h$ | Level |
| $V_{Cu}$, $Cu_{Zn}$ | $5\times10^{14}$ | (0/-)/acceptor | $10^{-13}$; $10^{-13}$ | 0.1eV above VBM Distribution VB tail |
| $V_{Sn}$,$Cu_{Sn}$, $Zn_{Sn}$ | $N_t$ $5\times10^{14}$ | Charge state; type (-/2-)/acceptor | $\sigma_e;\sigma_h$ $10^{-14}$;$10^{-14}$ | Level 0.2eV above VBM VB tail |
| $V_{Sn}$, $Cu_{Sn}$ | $N_t$ $5\times10^{14}$ | Charge state; type (2-/3-)/acceptor | $\sigma_e;\sigma_h$ $10^{-14}$;$10^{-14}$ | Level 0.5eV above VBM VB tail |
| $Zn_{Cu}$, $Sn_{Cu}$, $Sn_{Zn}$ | $N_t$ $5\times10^{14}$ | Charge state; type (3+/2+;2+/+;+/0)/donor | $\sigma_e;\sigma_h$ $10^{-13}$; $10^{-13}$ | Level 1.4eV above VBM CB tail |
| **Interface Defects** | | | | |
| Interface (CZTS/CdS) | Interface defect density (cm⁻²) $10^{12}$ | Charge state; type Neutral | $\sigma_e;\sigma_h$ $10^{-16}$;$10^{-18}$ | Uniform distribution |

*see the text in section 2 [54];. VBM (valence band maxima); VB tail (valence band tail); CB tail (Conduction band tail); unit of defect density ($N_t$) is cm⁻³ and that of electron and hole capture cross sections ($\sigma_e;\sigma_h$) is cm².

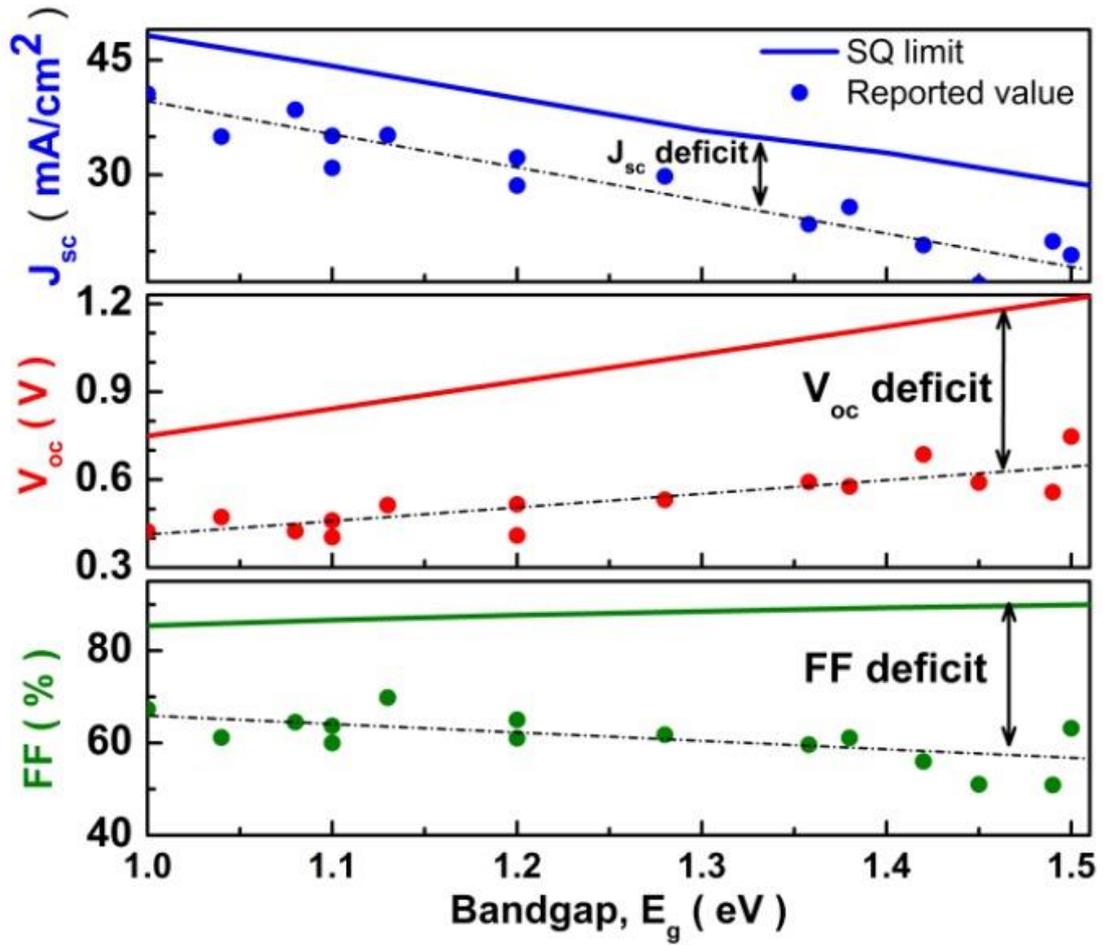

**Fig.1.**

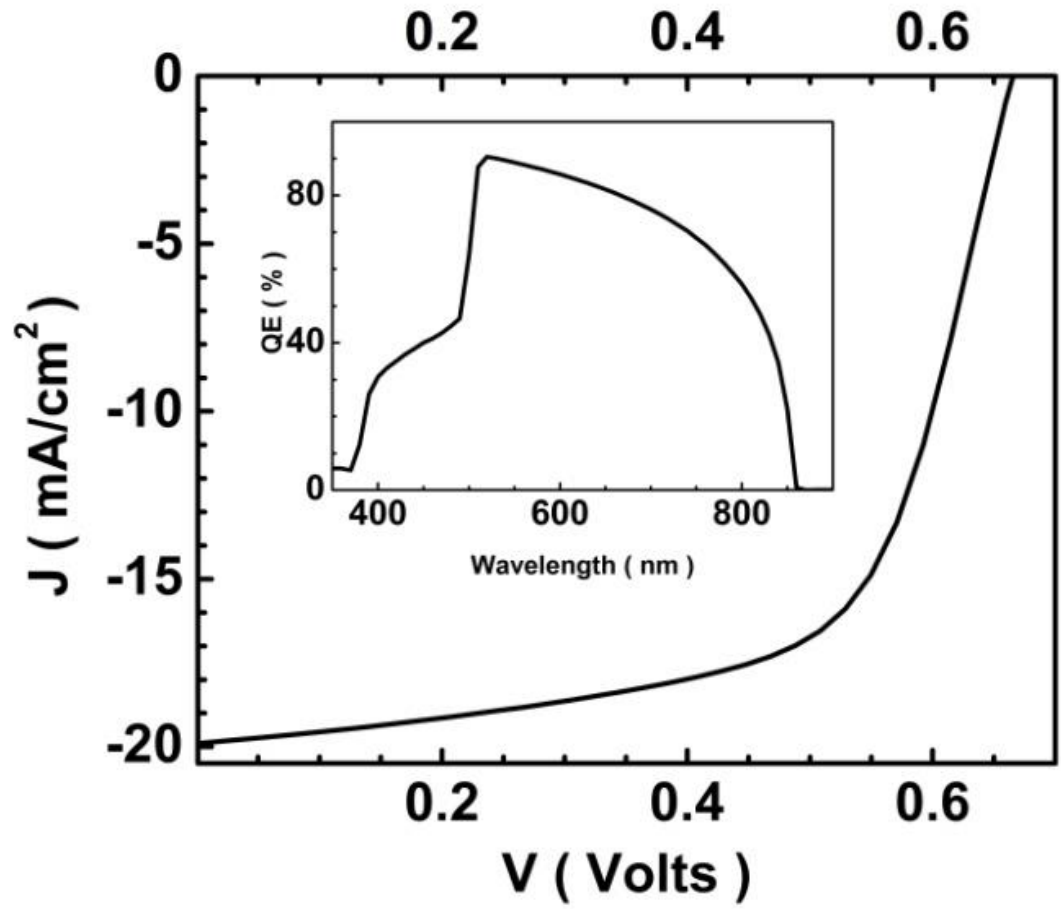

**Fig.2.**

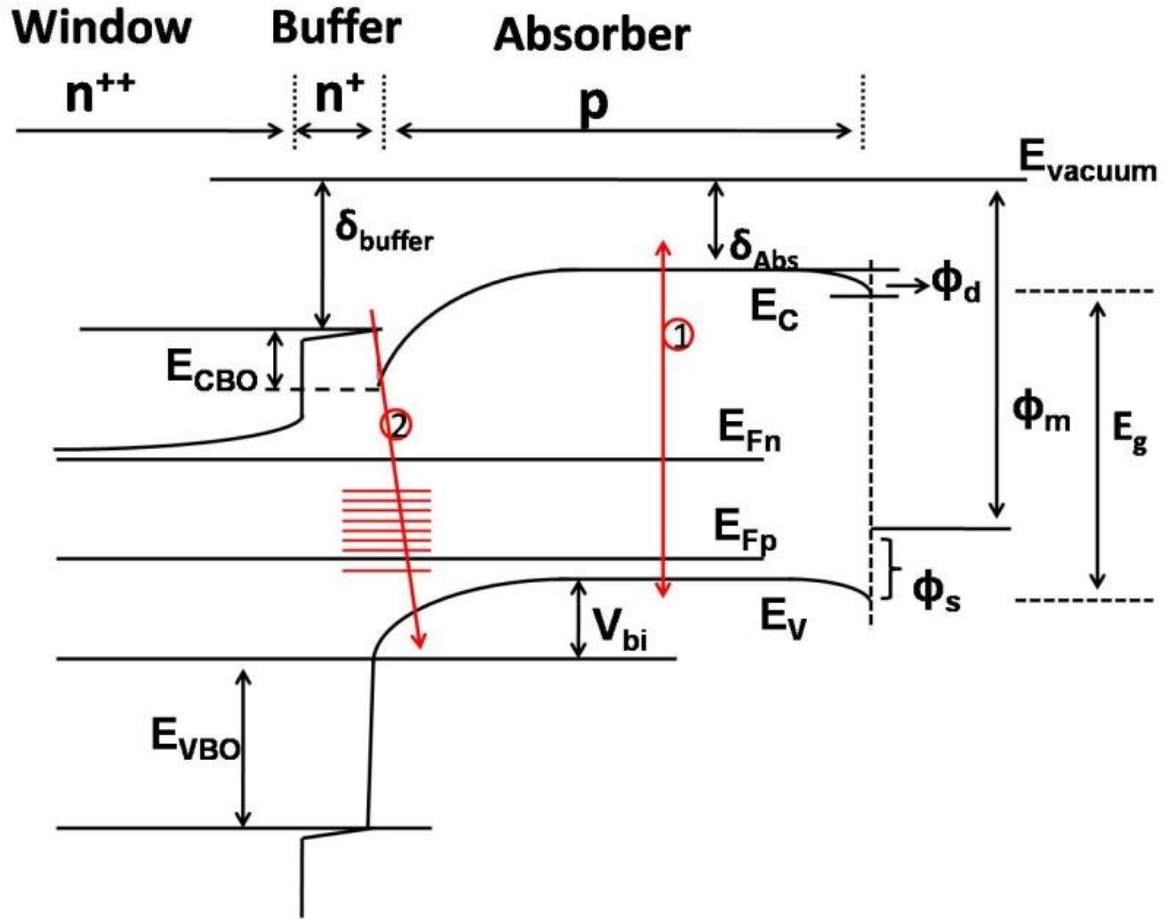

**Fig.3.**

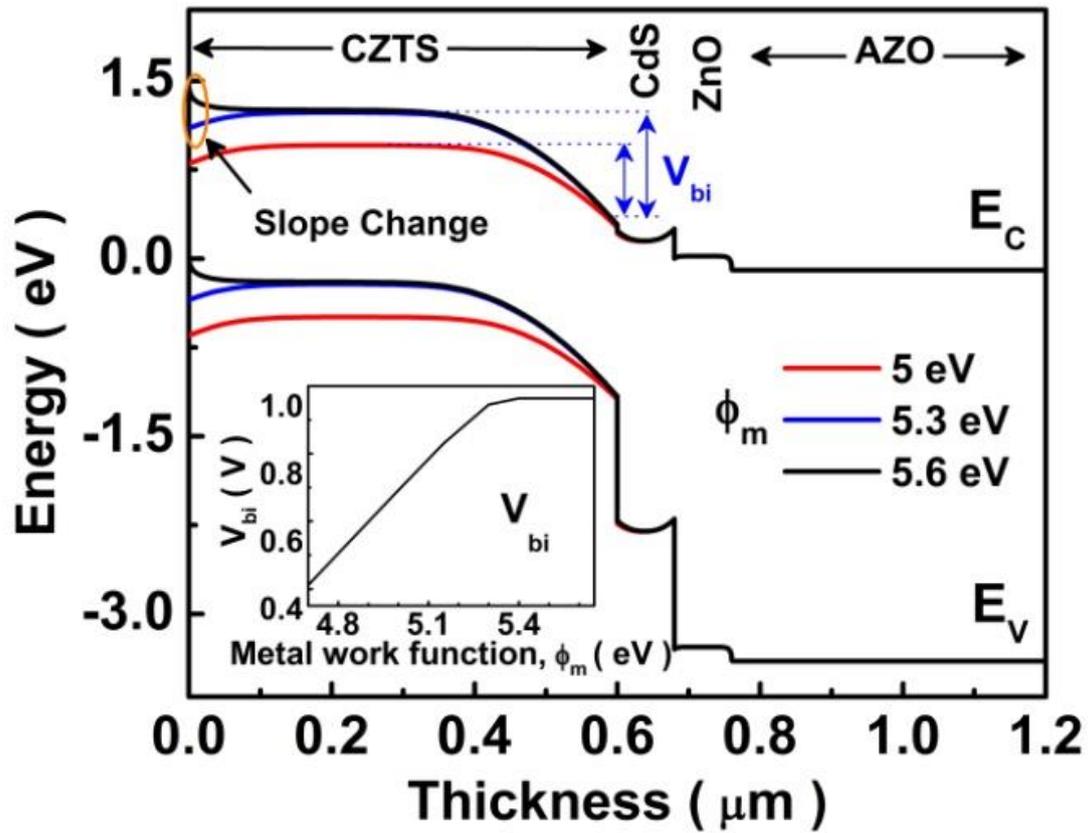

**Fig.4.**

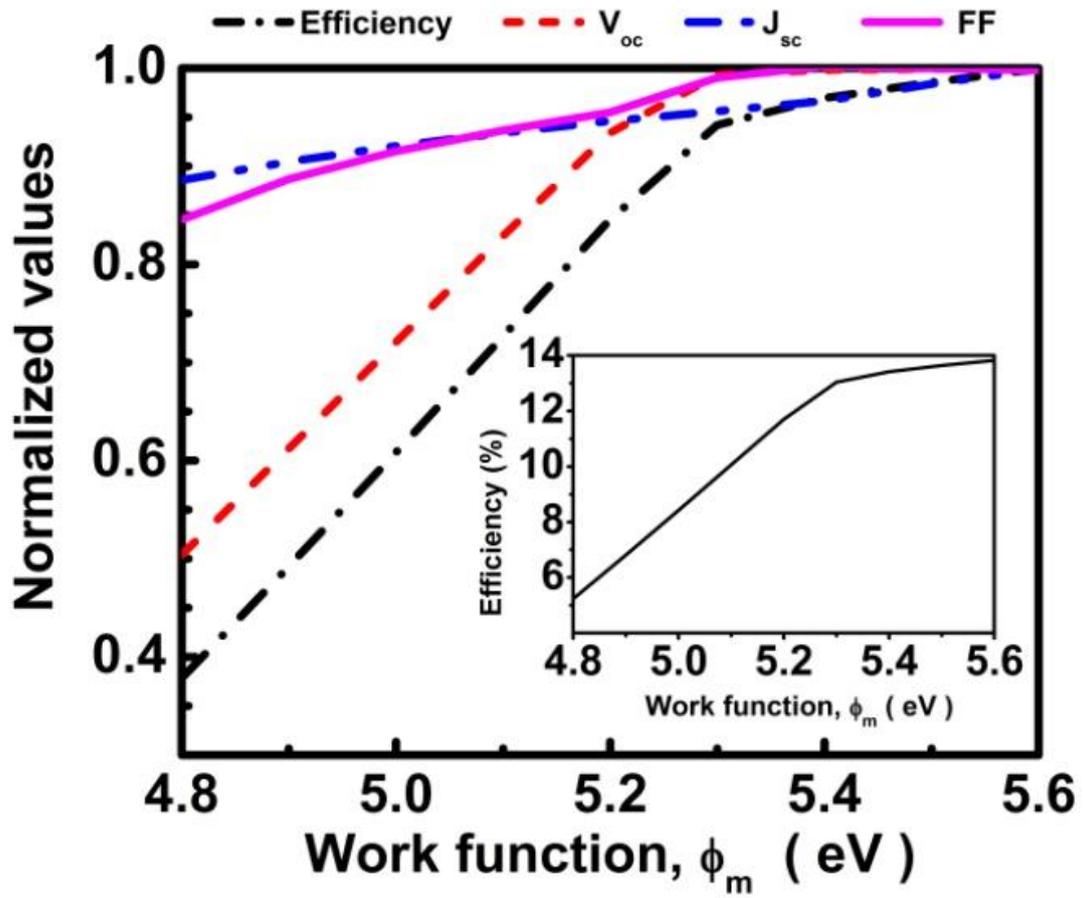

**Fig.5.**

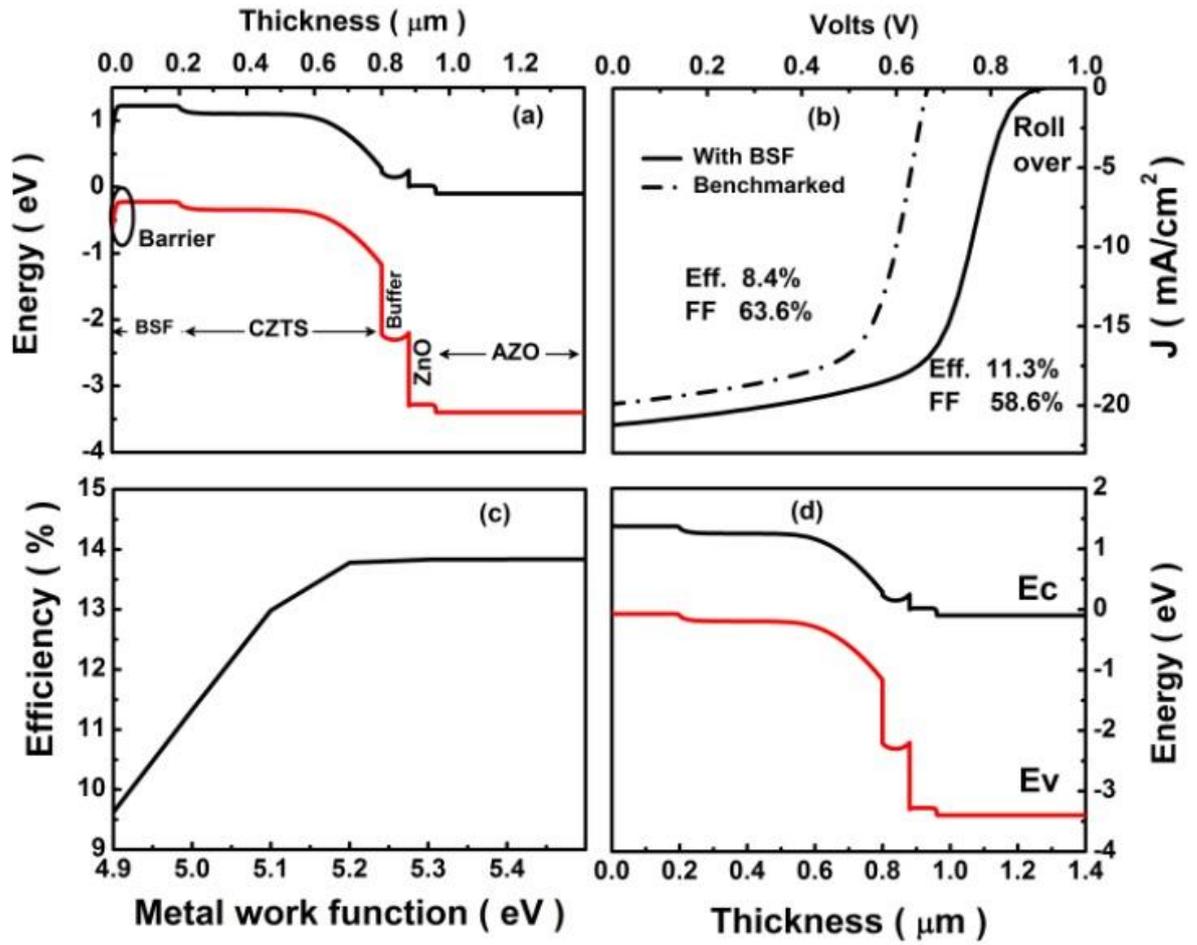

**Fig.6.**

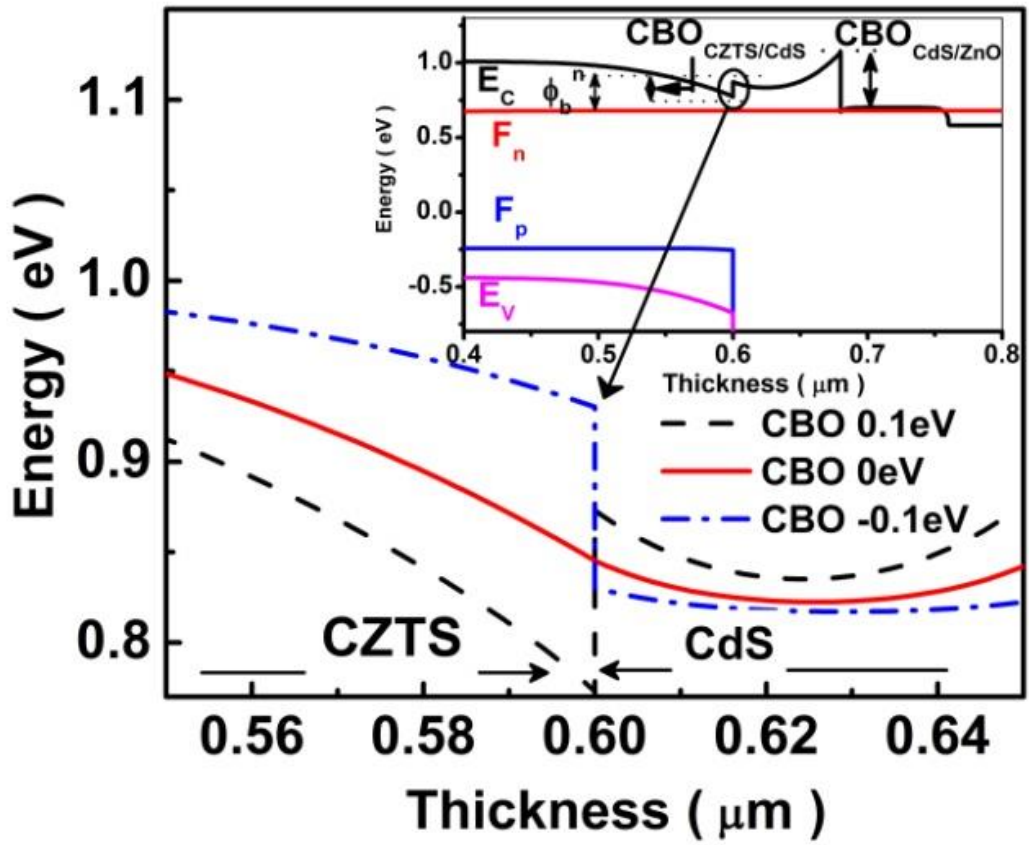

**Fig.7.**

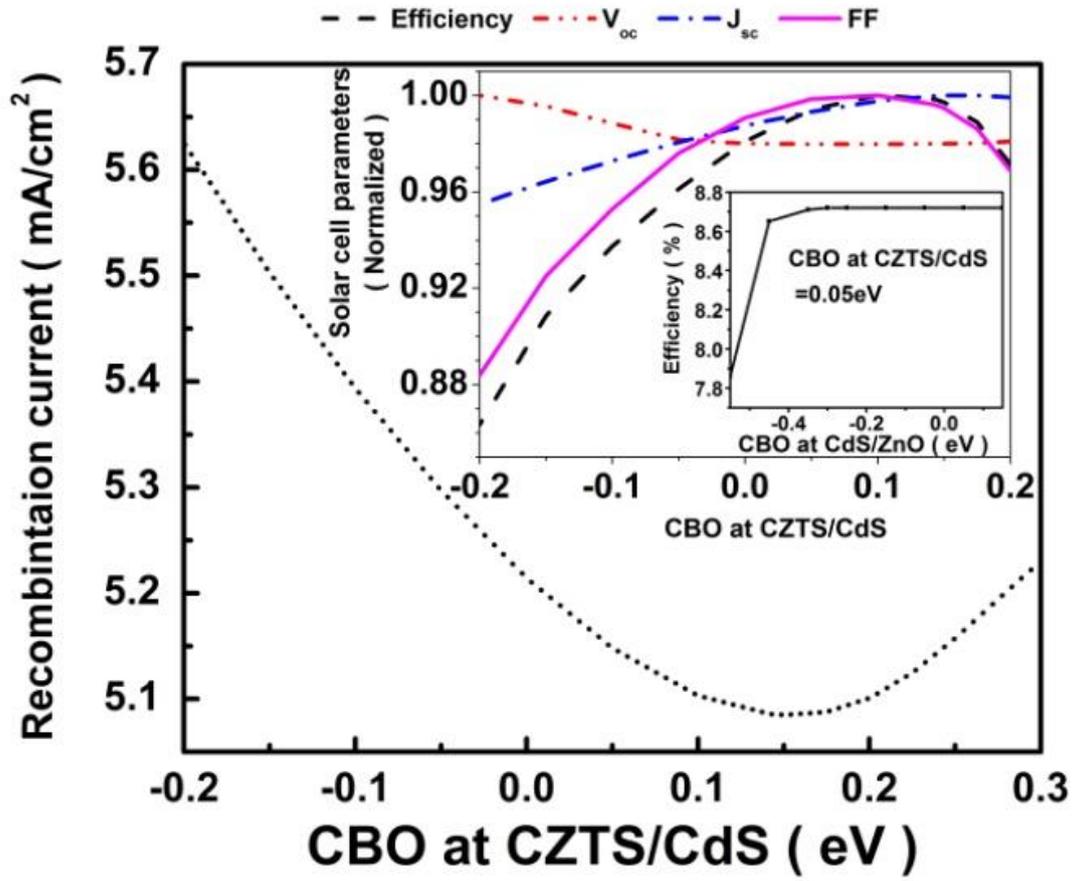

**Fig.8.**

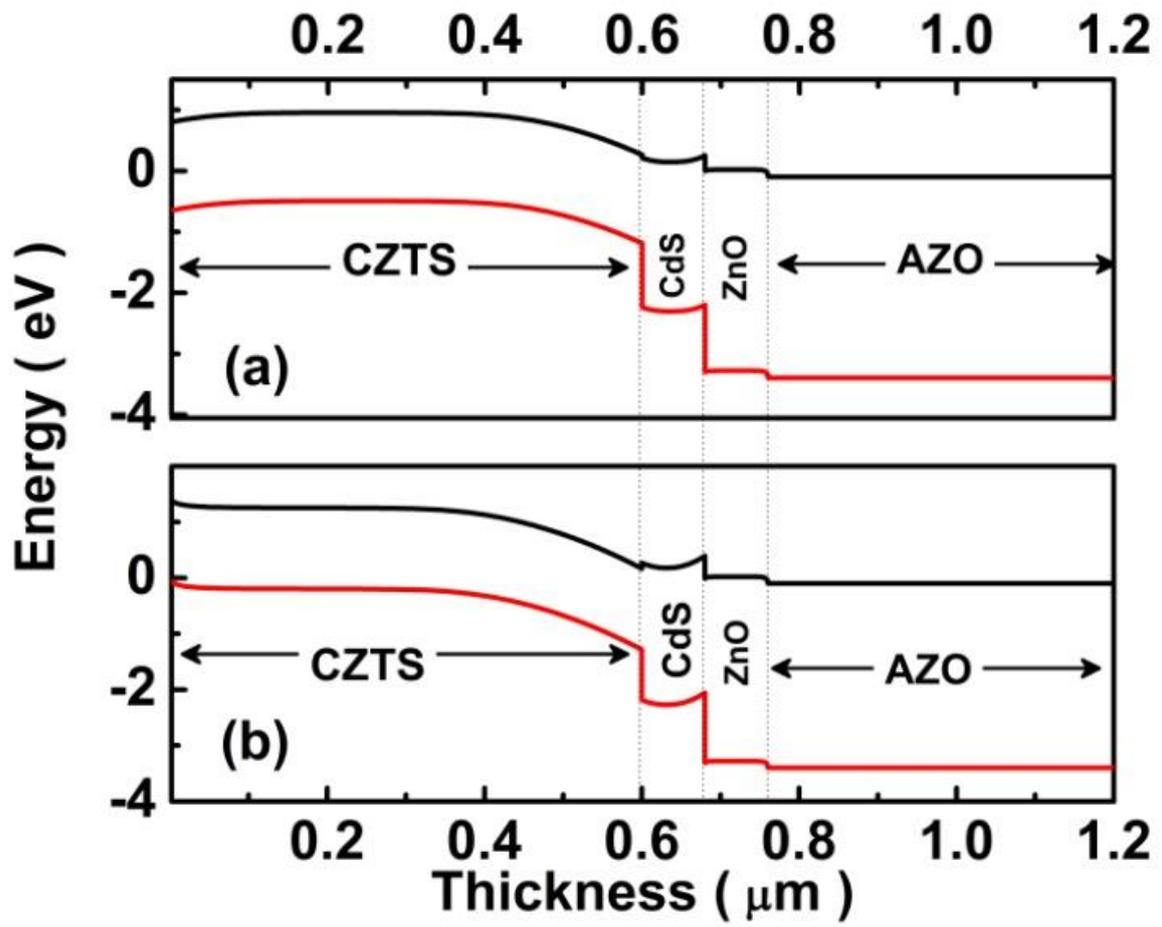

**Fig.9.**

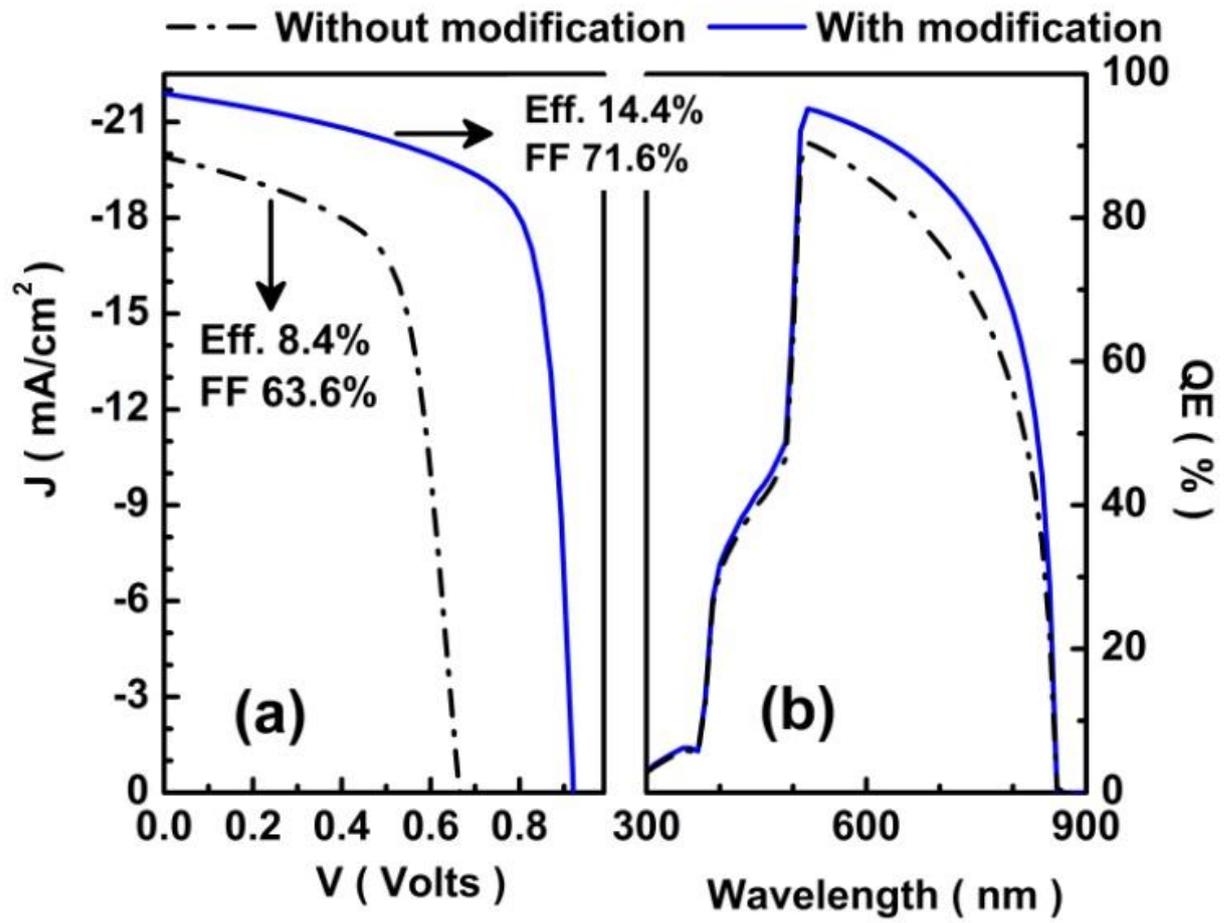

**Fig.10.**